\documentclass[times,twocolumn,review]{elsarticle}
\usepackage{medima}
\usepackage{framed,multirow}

\usepackage{amssymb}
\usepackage{latexsym}

\usepackage{url}
\usepackage{xcolor}

\usepackage{hyperref}
\usepackage{amsmath,amssymb,amsfonts}
\usepackage{algorithmic}
\usepackage{graphicx}
\usepackage{textcomp}
\usepackage{graphicx}
\usepackage{amsmath}
\usepackage{amssymb}
\usepackage{caption}
\usepackage{multirow}
\usepackage{booktabs}
\usepackage{makecell}
\usepackage{hyperref}
\usepackage{booktabs, tabularx}
\usepackage{soul}
\definecolor{newcolor}{rgb}{.8,.349,.1}

\journal{Medical Image Analysis}

\begin{document}

\verso{Chen Qin \textit{et~al.}}

\begin{frontmatter}

% \title{Biomechanics-informed generative model for myocardial motion tracking via latent space exploration}%
\title{Generative myocardial motion tracking via latent space exploration with biomechanics-informed prior}%
% \tnotetext[tnote1]{This is an example for title footnote coding.}

\author[1]{Chen \snm{Qin}\corref{cor1}}
% \cortext[cor1]{Corresponding author: e-mail: Chen.Qin@ed.ac.uk (Chen Qin)}
\author[2]{Shuo \snm{Wang}\corref{cor1}}
\cortext[cor1]{Corresponding authors: Chen.Qin@ed.ac.uk (Chen Qin);\\ shuowang@fudan.edu.cn (Shuo Wang)}
\author[3]{Chen \snm{Chen}}
%% Third author's email

\author[3,4]{Wenjia \snm{Bai}}
\author[3,5]{Daniel \snm{Rueckert}}

\address[1]{Institute for Digital Communications, School of Engineering, University of Edinburgh, Edinburgh, UK}
\address[2]{Digital Medical Research Center, School of Basic Medical Sciences, Fudan University, China}
\address[3]{Biomedical Image Analysis Group, Department of Computing, Imperial College London, London, UK.}
\address[4]{Department of Brain Sciences, Imperial College London, UK.}
\address[5]{Institute for AI and Informatics, Klinikum rechts der Isar, Technical University of Munich, Munich, Germany.}

\received{xx xx 2022}
\finalform{xx xx 2022}
\accepted{xx xx 2022}
\availableonline{xx xx 2022}
\communicated{xx xx}

\begin{abstract}
%%%
Myocardial motion and deformation are rich descriptors that characterize cardiac function. Image registration, as the most commonly used technique for myocardial motion tracking, is an ill-posed inverse problem which often requires prior assumptions on the solution space. In contrast to most existing approaches which impose explicit generic regularization such as smoothness, in this work we propose a novel method that can implicitly learn an application-specific biomechanics-informed prior and embed it into a neural network-parameterized transformation model. Particularly, the proposed method leverages a variational autoencoder-based generative model to learn a manifold for biomechanically plausible deformations. The motion tracking then can be performed via traversing the learnt manifold to search for the optimal transformations while considering the sequence information. The proposed method is validated on three public cardiac cine MRI datasets with comprehensive evaluations. The results demonstrate that the proposed method can outperform other approaches, yielding higher motion tracking accuracy with reasonable volume preservation and better generalizability to varying data distributions. It also enables better estimates of myocardial strains, which indicates the potential of the method in characterizing spatiotemporal signatures for understanding cardiovascular diseases.
%%%%
\end{abstract}


\end{frontmatter}

%\linenumbers

\section{Introduction}
Cardiac magnetic resonance imaging (MRI) is considered to be the clinical gold standard for non-invasive assessment of the morphology and function of the heart \citep{ismail2022cardiac}. It can provide important diagnostic information for cardiovascular diseases \citep{bello2019deep,zheng2019explainable}. Cardiac cine MRI is one of the most commonly used imaging modalities to capture the heart motion, which can reveal the contraction and relaxation pattern of the myocardium in fine spatial and temporal resolution. To explore the cardiac function, myocardial motion tracking is one of the crucial steps to enable the quantification of regional functions  \citep{duchateau2020machine}, such as changes in ventricular volumes and myocardial strains. However, manual quantification of myocardial motion can be hardly performed in clinical practice, and thus automatic approaches for accurate and reliable myocardial motion tracking are highly desirable.

Image registration techniques serve as an important tool for motion tracking analysis, which can infer anatomical correspondences between frames \citep{hajnal2001medical,qin2019machine}. Image registration is an ill-posed inverse problem, in which many viable solutions, i.e., spatial transformations, exist. To ensure the solution uniqueness and physiological plausibility, assumptions on the deformation fields are often imposed to convert the ill-posed problem to a well-posed one. To achieve this, most solutions propose to either parameterize the transformation model or add explicit regularization terms with desired properties \citep{rueckert2019model}, such as smoothness, diffeomorphism or incompressibility. Though these assumptions are generic, there is often a lack of incorporation of application-specific domain knowledge to inform the registration process. This may limit the transformation model in capturing more realistic and complex properties of moving organs which are not easy to be explicitly specified.
% Besides, the tuning of hyper-parameter $\lambda$ is non-trivial and often requires careful selection for specific registration problems.

On the other hand, data-driven strategies especially deep learning (DL)-based methods have become increasingly popular for image registration in recent years \citep{haskins2020deep}. 
The majority of methods either learn to regress the transformation parameters from a training set of registered images \citep{cao2017deformable,krebs2017robust,sokooti2017nonrigid}, or to minimize image dissimilarity metrics in a self-supervised way \citep{qin2018joint,balakrishnan2019voxelmorph}. 
These methods have shown competitive performance against traditional optimization-based approaches with faster registration speed. However, in contrast to conventional techniques which can well adapt to test data via optimizing the target alignments with respect to the transformation parameters, purely data-driven approaches inevitably suffer from the challenges in terms of the robustness and generalizability to domain shifts, such as variations in image acquisition parameters and protocols.

Motivated by these challenges, in this work, we propose a biomechanics-informed generative neural network for image registration via latent space exploration, with an application to left ventricle (LV) myocardial motion tracking in cardiac cine MRI. The proposed method provides an integrated data-driven and optimization model-based solution, which intends to incorporate the application-specific prior knowledge in a data-driven way and also to inherit the good generalizability of optimization-based models. Specifically, instead of imposing an explicit regularization term, the proposed model captures the prior knowledge about biomechanics by embedding it into an implicitly parameterized transformation model. This is realized via learning a  manifold for plausible deformations through a generative model, i.e., a temporal variational autoencoder (VAE), informed by domain-invariant biomechanical simulations. The decoder of VAE then serves as the transformation model that characterizes the biomechanical property and ensures the generation of plausible deformation fields from latent vectors. To adapt the learnt transformation function to myocardial motion tracking, latent space exploration is performed on the manifold while optimizing the dissimilarity metric on target sequences. This will enable the generated deformation to be regularized by the application-specific prior knowledge and will also guarantee good generalization capabilities in comparison to purely data-driven methods. 
%The method is evaluated in the context of myocardial motion tracking on cardiac cine MRI data. We show that the proposed method can achieve better performance compared to other competing approaches. It also indicates great potential in uncovering properties that are represented by biomechanics, with a more realistic range of clinical parameters such as radial and circumferential strains.

The main contributions of the work can be summarized as follows: First, we propose to learn a novel data-driven generative transformation model, which aims to implicitly capture the application-specific prior (e.g., biomechanical behavior of soft tissues) that might not be able to be explicitly specified. Second, we integrate the data-driven prior with latent space exploration as an optimization model-based image sequence registration, enabling more accurate and robust performance in comparison to learning-based approaches for cardiac MRI motion tracking. Furthermore, the proposed approach is data-efficient as it only relies on the domain-invariant biomechanical model during the training process. Compared with existing learning-based methods, it avoids the need for large training sets of imaging data.
Finally, extensive experiments were performed on three public cardiac MRI datasets, and we demonstrate that our approach is able to achieve better motion tracking performance against other competing methods in terms of accuracy, generalizability and estimation of down-stream clinical parameters. 
% This indicates its great potential in uncovering properties that are represented by the given data and its effectiveness in performing 

\section{Related Work}
Deformable image registration has been widely studied in the field of medical image analysis. The process of image registration can be defined as a minimisation problem, aiming to align between a source image $I_s$ and a target image $I_t$:
\begin{equation}
\label{objective_function}
\underset{\Phi \in \mathcal{D}(\Omega)}{\mathrm{argmin}}\  \mathcal{L}_{sim}(I_t, I_s \circ \Phi) + \lambda\mathcal{R}(\Phi).
\end{equation}
Here $\mathcal{L}_{sim}$ stands for the image dissimilarity measure, such as intensity changes or distances of anatomical landmarks; $\Phi$ denotes the transformation from $I_s$ to $I_t$; $\Omega$ is the image domain; $\mathcal{D}(\Omega)$ is the group of feasible transformations commonly parameterized by some basis functions (e.g. B-splines); $\mathcal{R}$ is a regularization term on the transformation field; and $\lambda$ is a hyper-parameter that trades-off between the image similarity and deformation regularity.

To incorporate the prior with desired properties for the deformation, the majority of methods either have built-in regularization constraints offered by the transformation model $\mathcal{D}(\Omega)$, or directly regularize the deformation field or its updates with $\mathcal{R}(\Phi)$ \citep{rueckert2019model}. The transformation model is one of the key aspects for image registration. Commonly, it can be compactly parameterized via spline-based models in a low dimensional representation space which inherently ensures the smoothness of the deformation \citep{rueckert1999nonrigid}, or densely parameterized at the pixel or voxel level such as in optical flow based methods \citep{1981Determining}.
% For instance, a rigid-body movement can be described by an affine transformation with six degrees of freedom including translations and rotations. For more complex deformations, the transformation $\mathcal{D}(\Omega)$ can also be compactly parameterised via spline-based models in a relatively low dimensional representation space \citep{rueckert1999nonrigid}, which inherently assumes the smoothness on the deformation. 
% Alternatively, deformations can also be densely parameterised at the pixel or voxel level, such as optical flow based methods \citep{1981Determining}. 
To consider the invertibility of the transformation, many diffeomorphic registration methods proposed to parameterize the displacement field as the integral of a time-varying or stationary velocity field \citep{beg05,ashburner07}. Other variants also considered enforcing volumetric preservation for soft tissue tracking \citep{fidon2019incompressible,mansi2011ilogdemons}. For explicit regularizers, most of the methods introduced regularization terms $\mathcal{R}(\Phi)$ such as in the form of L2 norm or bending energy to penalize the first- or second-order derivatives of the deformation \citep{sotiras2013deformable} to ensure the smoothness. Some other methods also proposed to constrain the optimization with some desired properties, such as imposing $\text{det}(J_{\Phi}(x))=1$ as a soft constraint (where $J_{\Phi}$ is the Jacobian matrix of the transformation $\Phi$) to guarantee tissue incompressiblity \citep{shi2012comprehensive}. 

%The majority of the methods explicitly assume the smoothness on the deformation, such as introducing a smoothness penalty in the form of L2 norm or bending energy to penalise the first- or second-order derivatives of the deformation \citep{sotiras2013deformable}, or parameterising the transformation via spline-based models in a relatively low dimensional representation space \citep{rueckert1999nonrigid}. To take into account the invertibility of the transformation, many diffeomorphic registration methods have been proposed to parameterise the displacement field as the integral of a time-varying velocity field or stationary velocity field (SVF) \citep{beg05,ashburner07}, which mathematically guarantees diffeomorphism. Volumetric preservation is also a characteriztic that is often desired for soft tissue tracking such as myocardium tracking to ensure that the total volume of myocardium keeps constant during image registration \citep{fidon2019incompressible,mansi2011ilogdemons,shi2012comprehensive}. Such incompressible registration methods either relax $\text{det}(J_{\Phi}(x))=1$ as a soft constraint \citep{shi2012comprehensive} (where $J_{\Phi}$ is the Jacobian matrix of the transformation $\Phi$), or use specific parameterisation for the displacement field or SVF \citep{fidon2019incompressible,mansi2011ilogdemons}. In addition, population statistics have also been investigated to incorporate prior knowledge for the registration process \citep{khallaghi2015statistical,rueckert2003automatic}. 

Recently, DL-based image registration has been shown to significantly improve the computational efficiency compared to conventional optimization-based methods. Similarly, most DL approaches regularize deformation fields with an explicit smoothness penalty \citep{Balakrishnan2018An,qin2018joint}, or parameterize transformations with free-form deformations using B-splines \citep{qiu2021learning} or with time-varying or stationary velocity fields \citep{balakrishnan2019voxelmorph,bone2019learning,krebs2019learning,qin2019unsupervised}. To leverage the  population-level statistics of the transformations, Bhalodia et al. \citep{bhalodia2019cooperative} has designed a population-based regularization to inform the registration to produce anatomically feasible correspondences. Besides, an adversarial learning approach for physically plausible deformation regularization was also proposed in a weakly-supervised anatomical-label-driven registration framework \citep{hu2018adversarial}. 
%In contrast to the existing deep learning based registration work, our method proposes to implicitly learn the regularization function via a biomechanics-informed VAE. This aims to capture deformation properties that are biomechanically plausible, with a particular focus on the application of myocardial motion tracking.

Myocardial motion tracking is one of the important applications for image registration. These methods commonly extended classical optical flow or image registration methods for cardiac motion estimation with a temporal regularization \citep{de2012temporal,shen2005consistent,shi2012comprehensive,tobon2013benchmarking}, such as using a series of free-form deformations \citep{shi2012comprehensive} or parameterizing a 4D velocity field with spatiotemporal kernels \citep{de2012temporal}. DL models have also been proposed for cardiac motion tracking \citep{qin2018joint,qin2018undersampled,krebs2019learning,2021Learning,qiu2019deep,lu2020going,zheng2019explainable,vos2019deep}. For instance, cardiac image segmentation and motion tracking were performed simultaneously in \citep{qin2018joint} while Krebs et al. \citep{krebs2019learning,2021Learning} proposed to learn a probabilistic motion model from sequences of cardiac MR images. However, most of the DL-based approaches only considered the generic spatiotemporal smoothness and did not take the physical plausibility of myocardial motion into account, which may adversely affect follow-up clinical parameter quantification such as strain estimation. Additionally, existing DL-based approaches rely on training with imaging data and do not normally consider test-time optimization, which may also limit their generalization potential.

Of particular relevance to this work, our previous work \citep{qin2020biomechanics} has proposed to learn a biomechanics-informed regularizer to capture deformation properties for myocardial motion tracking. Instead of learning an explicit regularization term for learning-based registration, this work proposes to implicitly parameterize the regularization into the transformation model for optimization-based registration, which allows the generated deformations to be physically plausible.
% and avoids the need for tuning hyper-parameter ($\lambda$). 
In addition, temporal information is introduced in this work via temporal VAE for the manifold learning and transformation model parameterization, and motion tracking is also performed with the temporal sequences taken into consideration. This has been shown to produce more consistent cardiac motion across time with better tissue volume preservation than independent pair-wise registration in our experiments. Significantly more thorough quantitative and qualitative evaluations of the proposed method including comparison, generalization and ablation studies have also been performed on a comprehensive set of cardiac cine MRI data.
%, and the registration is performed via optimizing the latent motion representation while traversing the biomechanically plausible deformation manifold. 

\section{Method}
\begin{figure*}[!t]
\centering
\includegraphics[width=\linewidth]{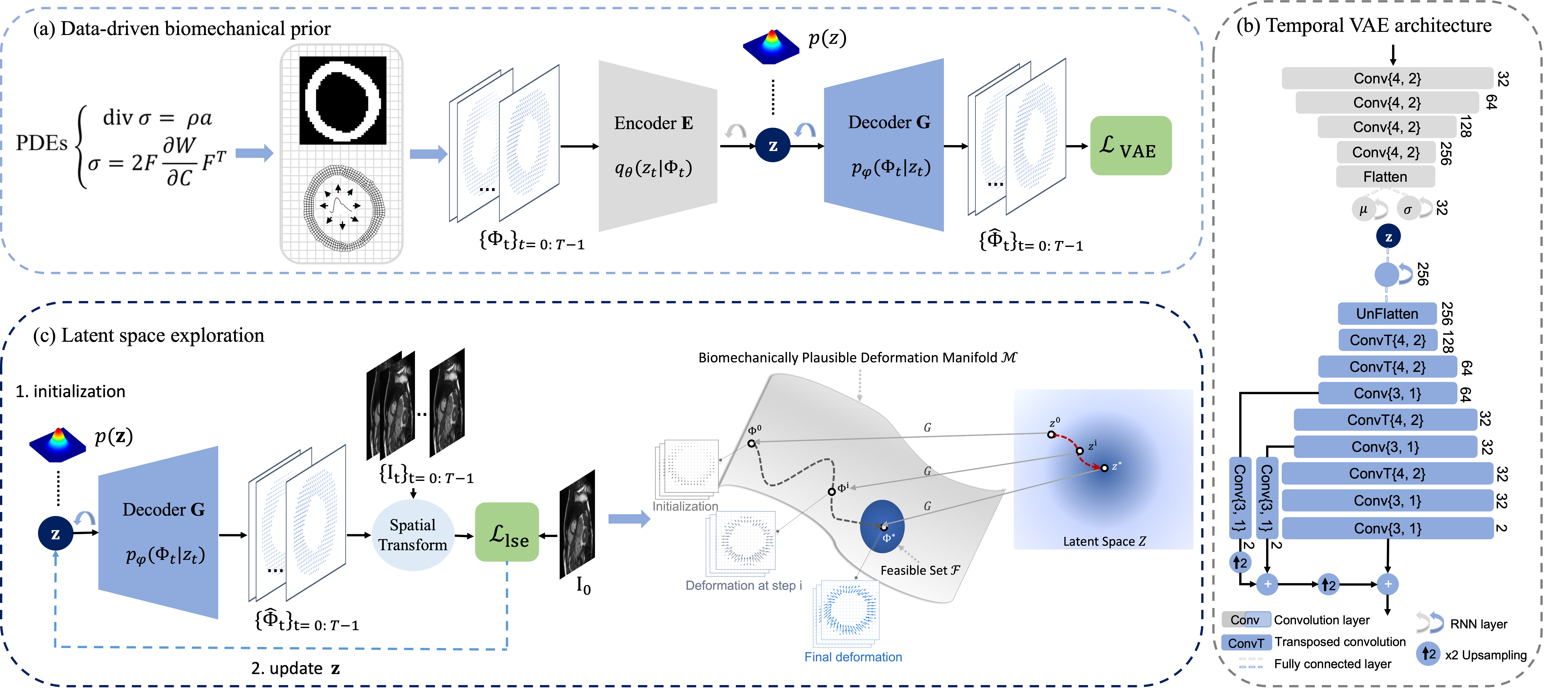}
\caption{Illustration of biomechanics-informed generative model for myocardial motion tracking: (a) The learning process of data-driven application-specific biomechanical prior; (b) The detailed network architecture for temporal VAE used in (a), where numbers in \{\} represent \{kernel size, stride size\} for 2D convolution; (c) Myocardial motion tracking via latent space exploration on the biomechanically plausible deformation manifold.}
\label{fig:framework}
\end{figure*}

The proposed biomechanics-informed registration model aims to generate physically plausible deformations for 2D myocardial motion tracking as well as to provide good generalizability. 
The proposed approach mainly consists of two parts: First, to capture the physical prior for the deformations, a plausible deformation manifold is learnt via a generative model based on domain-invariant biomechanical simulation; Then, the deformations can be compactly represented in a low dimensional latent space, where given unseen image sequences, the registration transformation can be optimized by traversing the plausible deformation manifold. An overview of the proposed framework is illustrated in Fig. \ref{fig:framework}. 

%The proposed biomechanics-regularized registration network is illustrated in Fig.\ref{fig:framework}, which mainly consists of three components. First, biomechanical simulations of deformation are generated according to equations which guarantee the physical properties of the deformation (Sec.~\ref{sec:biomechanical simulations}). Second, VAE is leveraged to learn the probability distribution of the simulated deformations, which implicitly captures the underlying biomechanical properties (Sec.~\ref{sec:VAE_reg}). Finally, the learnt VAE then acts as a regularization function for the registration network, which regularizes the solution space and helps to predict biomechanically plausible deformation without the need for any explicit penalty term (Sec.~\ref{sec:registration network}).

\subsection{Data-driven Application-specific Biomechanical Prior}
\subsubsection{Biomechanical simulations}
\label{sec:biomechanical simulations}
To capture an application-specific prior, we propose to learn a biomechanically plausible deformation manifold in a data-driven way, based on generating a set of deformation fields using computational simulation. From the perspective of solid mechanics, myocardial motion can be described with the following kinematic relations \citep{hunter1988analysis}:
\begin{equation}
{\rm div}\ \sigma = \rho a. 
\end{equation}
Here $\sigma$ is the stress tensor, ${\rm div}\ \sigma$ denotes the divergence of the stress tensor, $\rho$ is the material density and $a$ is the acceleration of cardiac muscle. The relationship between the deformation and stress is provided by the constitutive law, which is sophisticated due to both active and passive mechanical behaviours of cardiac muscle \citep{niederer2019short}. In this paper, we simplify the constitutive relations, assuming a hyper-elastic material behaviour:
\begin{equation}
\sigma = 2F  \frac{\partial W}{\partial C} F^T,
\end{equation}
where $F$ is the deformation gradient, $C$ is the right Cauchy–Green deformation tensor and $W$ stands for the strain energy density function (SEDF). We adopt a neo-Hookean constitutive model with the elastic modulus of 55.14 kPa from ex-vivo experiments~\citep{fatemifar2019comparison}. Although the anisotropic and electrophysiological characteristics of cardiac muscle microstructure are not included, this model is able to describe the incompressibility and large deformation of cardiac tissue. The above equations can be solved using the finite element method (FEM) once the geometry is reconstructed and boundary conditions are set.

We further simplify the cardiac deformation as a plane-strain problem considering the primary deformation within the short-axis plane. Therefore, a set of artificial deformation can be simulated based on slice-wise cardiac MR images and corresponding segmentations during a cardiac cycle. Firstly, the segmentation of myocardium $S_{ES}$ at the end-systolic (ES) frame is meshed with quadratic elements and treated as the computational start shape with zero pressure. Then the reference pressure $p_t$ for t-th frame is determined by minimising the area difference between the myocardium segmentation at t-th frame and the simulated pressurised shape from ES frame:
\begin{equation}
p_t = \mathop{\arg\min}\limits_{p} {||Area(S_{t})-Area(\Phi(S_{ES};p))||},
\end{equation}
where $S_{t}$ is the segmentation at t-th frame, $\Phi(\cdot;p)$ is the simulated deformation field under internal pressure $p$ and $Area$ represents the mathematical operation to calculate the size of a surface. Finally, an in-plane translation is added to the deformation field to align the center of deformed shape to the center of the real segmentation in each frame. Following the convention of cardiac motion tracking, we use the deformed shape at the end-diastolic (ED) frame as the reference configuration (Lagrangian description) and transform the deformation field accordingly. For t-th frame, the simulated deformation field $\Phi_t=\Phi(S_{ED};p_t)$ on the integration points are rasterized into image pixels of the MR images through bi-linear interpolation. 
% The codes for generating artificial deformation from cardiac segmentation are available at XXX (to be added).

\subsubsection{Biomechanically plausible deformation manifold}
\label{sec:VAE_reg}
Instead of explicitly parameterizing the mathematical formulation for the transformation model or specifying a regularization term, we propose to implicitly parameterize the transformation model informed by biomechnical prior knowledge. We achieve this by learning a biomechanically plausible deformation manifold with a VAE-based \citep{kingma2013auto} generative model, to learn the deep biomechanical prior via modelling the probability distribution of the plausible deformations. 
% To encode sequences of cardiac MR images, we propose a temporal VAE which considers the temporal information of the data and returns a distribution over the latent space. 
% which will enable the following latent optimization to move through the latent space continuously to determine a solution. 

Here we propose a temporal VAE with a recurrent multi-scale architecture to encode the sequence of deformation fields and return a distribution over the latent space, as shown in Fig. \ref{fig:framework}(b). Specifically, the network consists of an encoder which embeds the input, i.e. sequences of deformation fields from biomechanical simulation, onto a manifold represented by the latent space, and a multi-scale decoder that reconstructs the deformation from latent vectors sampled from the Gaussian distributions parameterized by estimated means $\mu$ and variances $\sigma$. To take into account of the temporal information of cardiac motion, recurrent units are incorporated into the VAE architecture to propagate the latent information along the temporal dimension, which aims to implicitly enforce the temporal coherence of the motion. {The recurrence is incorporated within both the encoder and decoder at the latent low-dimensional spaces through recurrent neural network (RNN) layers as shown in Fig. \ref{fig:framework}(b), to ensure that the estimated probability distributions of neighboring frames are temporally related and the reconstruction of deformation sequence can be informed of the temporal correlation. The recurrence over low-dimensional space also enables more efficient computation and optimization over temporal axis compared to the information propagation at higher dimensional feature space. Besides,} the multi-scale design of the decoder architecture, {shown as the skip and upsampling connections in the decoder}, also resembles conventional multi-resolution image registration framework, where multi-scale information can be exploited jointly for accurate alignment of images. 

% {\st{Note that this architecture is by no means optimal, however, as it is not the focus of this work, we do not study effects of the variations of architectures.}}

In detail, to learn the manifold for biomechanically plausible deformations, the temporal VAE model is trained to reconstruct the simulated sequence of deformations from Section \ref{sec:biomechanical simulations} as well as their first-order spatial derivatives {to alleviate effects of rigid
translation} \citep{qin2020biomechanics}. Let us denote the deformation between each image pair $(I_0, I_t)$ by {{$\Phi_t \in \mathrm{R}^{2\times M \times N}$}} in an image sequence of $I_{0:T-1}$, where $t\in [0, T-1]$, {$M$ and $N$ denote the spatial dimensions, $I_0$ (ED frame) is the target frame and $I_t$ is the source frame following the convention of cardiac motion tracking}. The corresponding first-order gradients of the deformation field can then be represented as $\triangledown \Phi_t \in \mathrm{R}^{4 \times M \times N}$. {In preliminary experiments \citep{qin2018joint,qin2020biomechanics}, we have empirically found that relatively large deformations (e.g., deformation between ED and ES frames) are more challenging to estimate compared to other frame pairs in a typical cardiac sequence. To alleviate this issue}, we propose to employ {a weighted {pixel-wise} reconstruction loss with the parameters $\mathbf{\omega}=\{\omega_0, \ldots, \omega_t, \ldots, \omega_{T-1}\}$ to put more weights on reconstructing large deformations} {during the training}, such that
\begin{equation}
\label{eq:VAE}
\begin{split}
        \mathcal{L}_{\mathrm{VAE}} = & \sum_{t=0}^{T-1} \omega_t \left[ \|\Phi_t - \hat{\Phi}_t\|_{2}^{2}+ \alpha \|\triangledown \Phi_t - \triangledown \hat{\Phi}_t\|_{2}^{2}\right]\\ &+ \frac{1}{T}\sum_{t=0}^{T-1} \beta \cdot D_{KL}(q_{\theta}(\mathbf{z}_t| {\Phi_t})\|p(\mathbf{z}_t)).
\end{split}
\end{equation}
The inputs to the VAE encoder are the sequences of deformation fields $\mathbf{\Phi}=\{\Phi_0, \ldots, \Phi_{T-1}\}$, and $\hat{\Phi}_t$ and $\triangledown \hat{\Phi}_t$ denote the VAE reconstruction from input $\Phi_t$ and its first-order spatial derivatives, respectively. {The loss function in Eq. \ref{eq:VAE} has taken into account the reconstruction of the deformation field along each axis. In addition,} we denote $\mathbf{z}\in \mathrm{R}^{D\times T}$ as the latent motion matrix {of a deformation sequence $\mathbf{\Phi}$ with $T$ frames} encoded by temporal VAE, and its column $\mathbf{z}_t\in \mathrm{R}^{D}$ corresponds to the latent encoding of deformation ${\Phi}_t$ at time step $t$. The manifold assumes a Gaussian prior on the latent variables $\mathbf{z}_t$, i.e. $p(\mathbf{z}_t) \sim \mathcal{N}(0,I)$. $q_{\theta}$ denotes the encoder parameterized by $\theta$ and $D_{KL}$ represents the Kullback-Leibler divergence. {Here $\omega$ is empirically set as weights drawn at fixed intervals from a 1D Gaussian window with peak value near ES frame for sequential time steps in a cardiac cycle} and are then normalized to $\sum_{t=0}^{T-1}\omega_t=1$. $\alpha$ is a hyper-parameter that balances between different reconstruction terms and $\beta$ controls the trade-off between reconstruction quality and the extent of latent space regularity. With a well-trained temporal VAE model, we expect that biomechanically plausible deformations can be generated from the latent space.

\subsection{Optimization Model-based Biomechanics-informed Motion Tracking}
\label{sec:registration network}
\subsubsection{Generative transformation model}
To perform the image registration, here we propose to leverage the pre-trained temporal VAE network as a generative transformation model. Particularly, we view the generation from the latent motion matrix $\mathbf{z}$ to the sequences of deformations, i.e., $\mathbf{\hat{\Phi}}=G(\mathbf{z})$, as the transformation model parameterized by $\varphi$, where $\varphi$ are pre-trained weights of the temporal VAE decoder. Given the biomechanically plausible deformation manifold as learnt in Section \ref{sec:VAE_reg}, the learning-based generative transformation model can implicitly embed the parameterization and regularization of the deformations into the neural network. This will enable $G$ to produce biomechanically plausible deformations without any explicit regularization. In addition, the training of the transformation model only requires biomechanical simulations which are generated based on segmentations. Different from other learning-based methods, it does not rely on any imaging data and is domain-invariant. Thus the learnt transformation model is less constrained to any particular cardiac dataset and can be more generalizable to multiple domains compared to standard DL-based approaches, as we will show in Section \ref{sec:results_quantitative} and \ref{sec:results_qualitative}.

\subsubsection{Motion tracking via latent space exploration}
{Given a sequence of images, the aim of latent space exploration for motion tracking is to traverse the manifold that is represented by the latent space of temporal VAE and search for the optimal transformation that can best register between them.} This is achieved via optimizing the image dissimilarity measure with respect to the latent motion matrix $\mathbf{z}$ during inference. 
As the ground truth dense correspondences between images are not available, the model thus learns to track the spatial features through time, relying on temporal intensity changes as self-supervision. The dissimilarity measure of the image sequence can be defined as
\begin{equation}
\label{L_sim}
   \mathcal{L}_{sim}(\mathbf{z})=\frac{1}{T}\sum_{t=0}^{T-1}\|(I_0-I_t \circ G(\mathbf{z}_t))\odot \mathbf{M}\|^2_2, 
\end{equation}
which is to minimize the pixel-wise mean squared error between the reference frame $I_0$ and transformed moving frames, {where $\circ$ represents the spatial transform, i.e., bilinear interpolation, as shown in Fig. \ref{fig:framework}(c)}. Here $\mathbf{M}$ is a binary dilated myocardial mask which is provided to enable the dissimilarity metric to focus on the region of interest, and $\odot$ represents the Hadamard product.  {Specifically, $\mathbf{M}$ is dilated from automatic myocardial segmentation at $I_0$, i.e., $S_0$, to cover the myocardial region as well as to take into account some neighboring background, which has been empirically shown to be able to better inform the alignment \citep{qin2020biomechanics}. Since we employ dilated masks for defining regions of interest, the accuracy of myocardial segmentation maps can thus be less required.} 
% Similarly, a weighted loss is also employed here to weigh more on more challenging registration.

To make sure that the generated deformations $G(\mathbf{z})$ fall on the learnt manifold while searching in the latent space, we also encourage the latent vectors to be in the region of high probability under the chosen Gaussian prior \citep{bora2017compressed}, by adding an $l_2$ regularization term in the loss function, such that
\begin{equation}
\label{L_sim_l2}
   \mathcal{L}_{lse}(\mathbf{z})=\frac{1}{T}\sum_{t=0}^{T-1}\|(I_0-I_t \circ G(\mathbf{z}_t))\odot \mathbf{M}\|^2_2+\mu \|\mathbf{z}\|^2_2. 
\end{equation}
Therefore the latent optimization will aim to minimize Eq.~\ref{L_sim_l2} while progressing along the given manifold of interest to guarantee the desired biomechanical plausibility, {where $\mu$ controls how likely $\mathbf{z}$ falls within the Gaussian prior}. 
% Note that here we do not use any explicit regularization term on the deformation along with the similarity metric, and thus it saves the effort for hyper-parameter tuning in comparison to conventional approaches. 
To optimize Eq. \ref{L_sim_l2}, Adam optimizer was employed to search for the latent representation $\mathbf{z}$ until the stop criterion $\mathcal{L}_{lse}^{i+1}-\mathcal{L}_{lse}^{i}\leq\epsilon$ was met, where $i$ represents the optimization iteration step. A detailed graphical illustration for the manifold traversing is shown in Fig. \ref{fig:framework}(c).

\section{Experiments}
\subsection{Data}
We evaluate the proposed method on three cardiac cine MRI datasets. 

\subsubsection{UK Biobank dataset} The first dataset consists of 300 short-axis cardiac cine MRI sequences from the publicly available UK Biobank (UKBB) dataset \citep{petersen2017reference}. Each sequence consists of 50 frames and each frame forms a 2D stack of typically 10 image slices, with the spatial resolution of 1.8$\times$1.8$\times$10 mm. Segmentation masks for LV myocardium were obtained via an automated tool provided in \citep{bai2018automated}. Data used for biomechanical simulations were taken from a separate subset in UKBB cohort consisting of 200 healthy subjects. 
\subsubsection{ACDC dataset} The Automated Cardiac Diagnosis Challenge (ACDC) dataset \citep{bernard2018deep}\footnote{{https://www.creatis.insa-lyon.fr/Challenge/acdc/databases.html}} is a public dataset which consists of 100 cardiac cine MRI sequences with 5 evenly distributed subgroups (4 pathological plus 1 healthy subject groups). A series of short-axis slices cover the LV from the base to the apex, with one slice every 5 or 10 mm according to the examination. The in-plane resolution goes from 1.37 to 1.68 mm. The temporal sequences consist of frames with numbers ranging from 12 to 35. The LV myocardium were manually labeled by experts.
\subsubsection{M\&Ms dataset} We used 150 cardiac cine MR sequences from two different MRI vendors (75 each) from the Multi-Centre, Multi-Vendor \& Multi-Disease Cardiac Image Segmentation Challenge (M\&Ms) dataset \citep{campello2021multi}\footnote{{https://www.ub.edu/mnms/}}. The in-plane resolution ranges from 0.98 to 1.45 mm and the slice thickness goes from 9.2 to 9.9 mm. The used sequences consist of either 25 or 30 frames. The LV myocardium has been segmented by experienced clinicians from the respective institutions.

\subsection{Evaluation Method}
\label{sec:evaluation_method}
We compared our proposed approach with representative open-sourced motion estimation methods, including classical conventional approaches such as free-form deformation (FFD) with volumetric preservation (VP) \citep{rohlfing2001intensity} and diffeomorphic Demons (dDemons) \citep{vercauteren2007non}, as well as two state-of-the-art DL-based approaches, i.e., Motion-Net \citep{qin2018joint} and biomechanics-informed neural network (BINN) \citep{qin2020biomechanics}. In contrast to our proposed approach in this work, Motion-Net \citep{qin2018joint} employs an explicit regularizer in the form of an approximation of Huber loss to penalize displacement gradients and BINN \citep{qin2020biomechanics} is a counterpart that learns an explicit regularization term for biomechanical properties. Specifically, the compared DL baselines were trained on 2D stacks of UKBB data with 100/50 training/validation data, and were tested on the remaining 150 test data. 
% Data used for biomechanical simulations were from a separate data set in UKBB consisting of 200 subjects. 
For fair comparisons, reference masks $\mathbf{M}$ were also employed to define the domain within which to evaluate the energy function for FFD-VP and BINN approaches. 

Quantitative results were evaluated in terms of mean contour distance (MCD), Dice coefficient, and the mean absolute difference between Jacobian determinant $\text{det}(J_{\Phi}(x))$ and 1 over LV myocardium, denoted as $||J|-1|$ for simplicity. These metrics were made to measure the motion estimation performance with complimentary emphasis, {for instance, Dice coefficient and MCD measure the region overlap and averaged contour alignment reflecting registration accuracy and $||J|-1|$ evaluates the level of volume preservation. The higher the Dice value and the lower the MCD and $||J|-1|$ indicate the better performance.} Besides, the performance of the proposed and baseline methods were also compared in terms of LV myocardial strain estimates. These were all used as means to evaluate the learnt biomechanical property. The strains were calculated on cine MRI using the Lagrangian strain tensor formula \citep{elen2008three} and the peak strains were computed between ES and ED frames. 
% In experiments, these evaluations were performed on different representative slices, i.e., apical, mid-ventricle, and basal slices separately. 
Statistical test {of Wilcoxon signed-rank test \citep{wilcoxon1992individual}} was also performed for results of each metric to test if the differences between methods were significant. 
% {We first performed a Friedman test \citep{friedman1937use} to see if there was a significant difference in the metric results. Then if the null hypothesis of the Friedman Test was rejected, we performed one-versus-all one-way Wilcoxon signed-rank test \citep{wilcoxon1992individual} with Bonferroni correction to find out if our model significantly outperformed the others.}

\subsection{Implementation Details}
The detailed temporal VAE architecture is described in Fig. \ref{fig:framework}(b). The latent vector dimension $D$ was set to 32 for each $\mathbf{z}_t$ {based on our ablation study in Section \ref{sec:varying_z}}, and $\alpha=10$ and $\beta = 0.01$ in Eq. \ref{eq:VAE} {were determined based on the validation set}. $\mathbf{\omega}$ was empirically set as values sampled from $\mathcal{N}(60,10)$ with fixed intervals starting at 0 for $T=50$ steps, {where the first 20 frames were sampled between [0, 60] with an interval of 3 and the last 30 frames were sampled from [60, 120] with an interval of 2, as we empirically found that ES frames in our sequences were mostly near $T=20$}.
A learning rate of 0.0001 with Adam optimizer was employed to train the temporal VAE for 200 epochs. For the latent space exploration, $\mu$ in Eq. \ref{L_sim_l2} was set to 0.001 and a learning rate of 0.1 with Adam optimizer was used to optimize the latent representation $\mathbf{z}$ for motion tracking, where $\mathbf{z}$ was initialized with random sampling from $\mathcal{N}(0, 0.8)$. Additionally, all the images were resampled to a spatial resolution of $1.8\times 1.8$ mm, and image intensity was linearly normalized to the range of $[0,1]$. Hyper-parameters in all the comparison methods were selected to ensure that they can achieve an average of $||J|-1|$ near 0.1. The proposed method and the compared DL baselines were all implemented in PyTorch, and experiments were performed on a 12GB Nvidia Titan Xp Graphics Processing Unit (GPU). For FFD-VP and dDemons, experiments were conducted on a 16GB RAM, 3.60GHz Central Processing Unit (CPU). The biomechanical simulation is solved with the FEM commercial software (ADINA R\&D Inc.).  Source code will be made available online upon publication\footnote{{https://github.com/cq615/BIGM-motion-tracking}}.

\subsection{Results}
\subsubsection{Quantitative comparison}
\label{sec:results_quantitative}
We first evaluated the method comparisons and their generalization capabilities on the given three cardiac cine MRI datasets. Quantitative comparisons were performed on three representative image slices, namely a basal slice at 25\% of the LV length, a mid-ventricle slice at 50\% and an apical slice at 75\% of the LV length, according to \citep{taylor2015myocardial,schuster2015cardiovascular}. {Evaluations on these representative slices can provide better understanding of the motion tracking performance at different planes of the cardiac volume than averaging over all image slices.}
Test results on the three different datasets are presented in Table \ref{motion_evaluation}, {where it shows the registration performance between ED and warped ES frames, i.e., the largest deformation in a sequence}. Bold numbers indicate the best performance and the underlined results represent the second best performance among all the competing methods. On UKBB dataset, it can be observed that the proposed method can achieve competitive performance on all these slices in terms of both registration accuracy (MCD and Dice) and volume preservation ($||J|-1|$) compared with other methods. Most of the best and second best results fall within the approaches which are biomechanics-informed (BINN and the Proposed), demonstrating the benefits of incorporating the application-specific prior knowledge. The performance gains are especially significant on more challenging slices such as apical and basal slices, where prior knowledge of physically plausible deformation can have more impact on informing the alignment. It is worth to note that the DL-based baselines (Motion-Net and BINN) were trained on UKBB dataset which utilised the in-distribution data to learn the registration networks, whereas the proposed approach did not take advantage of such imaging information during training.

We have also performed the quantitative comparisons on two other datasets, ACDC and M\&Ms, as a \textbf{generalization study} of the approaches. {Since recurrent units can deal with arbitrary sequence length, our proposed approach can well adapt to temporal sequences with different number of frames.} For Motion-Net and BINN, we deployed their models trained on UKBB dataset and directly tested them on 100 ACDC data and 150 M\&Ms data. The test results are also shown in Table \ref{motion_evaluation}. Similar to the results on UKBB dataset, most of the top performances were achieved on methods that are biomechanically informed. 
% This is likely due to the benefit of the biomechanical constraint either in explicit or implicit form, which enforces the generated deformations to be biomechanically plausible and less sensitive to domain shift problem in the learning setting. 
In terms of the generalization performance, a significant improvement of the proposed method can be observed against other baselines, including both the conventional optimization-based approaches and the DL-based methods. Though BINN and the proposed method achieved comparable performance on UKBB data, here the proposed method generalized better on data in unseen domains (ACDC and M\&Ms), with a higher registration accuracy, i.e., an average improvement of 2.4\% in terms of Dice and 0.28 mm in terms of MCD by averaging the results from both datasets (statistically significant with $p \ll 0.001$). We will also discuss these results in Section \ref{sec:discussion}.
% This can be partially explained by the fact that the proposed method is imaging domain-invariant which does not encode any imaging information, and thus will not be affected by imaging data distribution shift. 
% Apart from that, the latent space exploration on the biomechanically plausible manifold also offers the flexibility in searching for the optimal transformation for registration, which can be effectively guided by the similarity measure in target sequences during inference.

\begin{table*}[!t]
  \centering
  \caption{Quantitative comparisons of myocardial motion estimation performance among FFD-VP~\citep{rohlfing2001intensity}, dDemons \citep{vercauteren2007non}, Motion-Net \citep{qin2018joint}, BINN \citep{qin2020biomechanics} and the proposed method. Results {are measured between ED and warped ES frames and} are reported as mean (standard deviation). Lower MCD (unit: mm) and higher Dice indicates better accuracy. Lower $||J|-1|$ indicates better volume preservation. \textbf{Bold} indicates the best performance and the \underline{underline} indicates the second best results.}
  \label{motion_evaluation}
\scalebox{0.75}
% \scriptsize
 { \begin{tabular}{ccccccccccc}
  
    \toprule
\multirow{2}*{Dataset}& \multirow{2}*{Method} & \multicolumn{3}{c}{Apical} & \multicolumn{3}{c}{Mid-ventricle}& \multicolumn{3}{c}{Basal}\\
% \cline{3-11}
\cmidrule{3-11}
 & & {MCD} & {Dice}   & $||J|-1|$  & {MCD} & {Dice}   & $||J|-1|$   & {MCD} & {Dice}   & $||J|-1|$  \\
  \midrule
%   \multirow{5}*{UKBB}& {FFD-VP \citep{rohlfing2001intensity}} & \makecell{2.209(1.870)} &\makecell{0.622(0.183)} &\makecell{0.174(0.074)} & \makecell{\underline{1.362}(0.447)}&\makecell{0.781(0.082)}&\makecell{0.176(0.037)} &\makecell{2.360(1.613)} &\makecell{0.680(0.192)} &\makecell{0.198(0.067)}\\

 \multirow{5}*{UKBB}& {FFD-VP} & \makecell{2.529(1.291)} &\makecell{0.622(0.196)} &\makecell{0.195(0.062)} & \makecell{{2.130}(0.734)}&\makecell{0.733(0.078)}&\makecell{0.178(0.046)} &\makecell{4.184(2.104)} &\makecell{0.529(0.210)} &\makecell{0.174(0.040)}\\

%   \midrule
  
 &{dDemons} & \makecell{2.178(1.079)} & \makecell{0.651(0.186)} & \makecell{0.133(0.044)} & \makecell{1.603(0.629)} & \makecell{0.784(0.072)} & \makecell{0.148(0.036)} & \makecell{3.568(1.482)} & \makecell{0.560(0.176)} & \makecell{\underline{0.132}(0.029)}\\
%  \midrule
&Motion-Net & \makecell{2.177(1.119)}&\makecell{0.639(0.196)}&\makecell{\textbf{0.087}(0.028)}&\makecell{{1.442}(0.625)}&\makecell{\textbf{0.806}(0.063)}&\makecell{\underline{0.104}(0.042)}&\makecell{2.912(1.384)}&\makecell{0.619(0.178)}&\makecell{0.149(0.047)} \\
% \midrule
&\makecell{BINN} & \makecell{\textbf{1.734}(0.808)}&\makecell{\underline{0.684}(0.160)} &\makecell{\underline{0.126}(0.071)}&\makecell{\underline{1.417}(0.402)} & \makecell{0.796(0.059)} &\makecell{\textbf{0.081}(0.020)} & \makecell{\underline{2.280}(1.274)} & \makecell{\textbf{0.725}(0.149)} &\makecell{\textbf{0.124}(0.037)}\\
% \midrule
% \midrule
&\makecell{Proposed} & \makecell{\underline{1.776}(0.974)}&\makecell{\textbf{0.689}(0.145)} &\makecell{0.139(0.091)}&\makecell{\textbf{1.321}(0.403)} & \makecell{\underline{0.805}(0.064)} &\makecell{{0.121}(0.039)} & \makecell{\textbf{2.246}(1.430)} & \makecell{\underline{0.715}(0.147)} &\makecell{{0.159}(0.062)}\\
\midrule
  \multirow{5}*{ACDC}& {FFD-VP} & 2.784(1.648) &0.684(0.143)&0.148(0.075) &2.442(1.352) &0.756(0.088)&0.147(0.072) &2.925(1.328) &0.731(0.120) &\underline{0.135}(0.052)\\
%   \midrule
  
 &{dDemons} & \makecell{2.437(1.453)} & \makecell{\underline{0.707}(0.128)} & \makecell{\textbf{0.132}(0.043)} & \makecell{\underline{1.993}(1.112)} & \makecell{\underline{0.788}(0.075)} & \makecell{\underline{0.141}(0.048)} & \makecell{2.295(1.174)} & \makecell{0.778(0.102)} & \makecell{\textbf{0.132}(0.031)}\\
%  \midrule
&Motion-Net & \makecell{2.853(1.301)}&\makecell{0.656(0.142)}&\makecell{{0.171}(0.060)}&\makecell{{2.799}(1.007)}&\makecell{{0.745}(0.105)}&\makecell{0.167(0.056)}&\makecell{2.814(1.236)}&\makecell{0.751(0.123)}&\makecell{0.167(0.049)} \\
% \midrule
&\makecell{BINN} & \makecell{\underline{2.450}(1.253)}&\makecell{\underline{0.707}(0.147)} &\makecell{0.158(0.081)}&\makecell{{2.210}(0.918)} & \makecell{0.783(0.097)} &\makecell{{0.148}(0.056)} & \makecell{\underline{2.229}(0.860)} & \makecell{\underline{0.789}(0.091)} &\makecell{{0.161}(0.063)}\\
% \midrule
% \midrule
&\makecell{Proposed} & \makecell{\textbf{2.101}(1.312)}&\makecell{\textbf{0.731}(0.137)} &\makecell{\underline{0.134}(0.091)}&\makecell{\textbf{1.646}(0.971)} & \makecell{\textbf{0.818}(0.055)} &\makecell{\textbf{0.134}(0.067)} & \makecell{\textbf{1.660}(0.671)} & \makecell{\textbf{0.829}(0.067)} &\makecell{{0.138}(0.060)}\\
\midrule
  \multirow{5}*{M\&Ms}& {FFD-VP} &3.469(1.917)  & 0.612(0.197)&0.175(0.069) & 2.589(1.228)&0.729(0.112)&0.136(0.050)&2.534(1.323) &0.695(0.165) &0.177(0.084)\\
%   \midrule
  
 &{dDemons} & \makecell{3.045(1.738)} & \makecell{0.640(0.190)} & \makecell{\textbf{0.153}(0.042)} & \makecell{1.995(0.918)} & \makecell{0.772(0.091)} & \makecell{{0.143}(0.039)} & \makecell{1.991(1.093)} & \makecell{0.737(0.155)} & \makecell{{0.166}(0.051)}\\
%  \midrule
&Motion-Net & \makecell{3.174(1.574)}&\makecell{0.604(0.187)}&\makecell{{0.198}(0.057)}&\makecell{{2.092}(0.981)}&\makecell{{0.748}(0.103)}&\makecell{0.151(0.078)}&\makecell{2.209(1.196)}&\makecell{0.715(0.122)}&\makecell{\underline{0.156}(0.080)} \\
% \midrule
&\makecell{BINN} & \makecell{\underline{2.577}(1.261)}&\makecell{\underline{0.687}(0.152)} &\makecell{0.174(0.074)}&\makecell{\underline{1.826}(0.658)} & \makecell{\underline{0.790}(0.087)} &\makecell{\underline{0.131}(0.050)} & \makecell{\underline{1.881}(0.828)} & \makecell{\underline{0.752}(0.094)} &\makecell{\underline{0.156}(0.089)}\\
% \midrule
% \midrule
&\makecell{Proposed} & \makecell{\textbf{2.498}(1.401)}&\makecell{\textbf{0.705}(0.149)} &\makecell{\underline{0.155}(0.065)}&\makecell{\textbf{1.719}(0.903)} & \makecell{\textbf{0.814}(0.068)} &\makecell{\textbf{0.124}(0.052)} & \makecell{\textbf{1.875}(1.106)} & \makecell{\textbf{0.759}(0.104)} &\makecell{\textbf{0.152}(1.106)}\\
    \bottomrule
  \end{tabular}}
\end{table*}

\subsubsection{Qualitative comparison}
\label{sec:results_qualitative}
\begin{figure*}[!t]
\centering
\includegraphics[width=\linewidth]{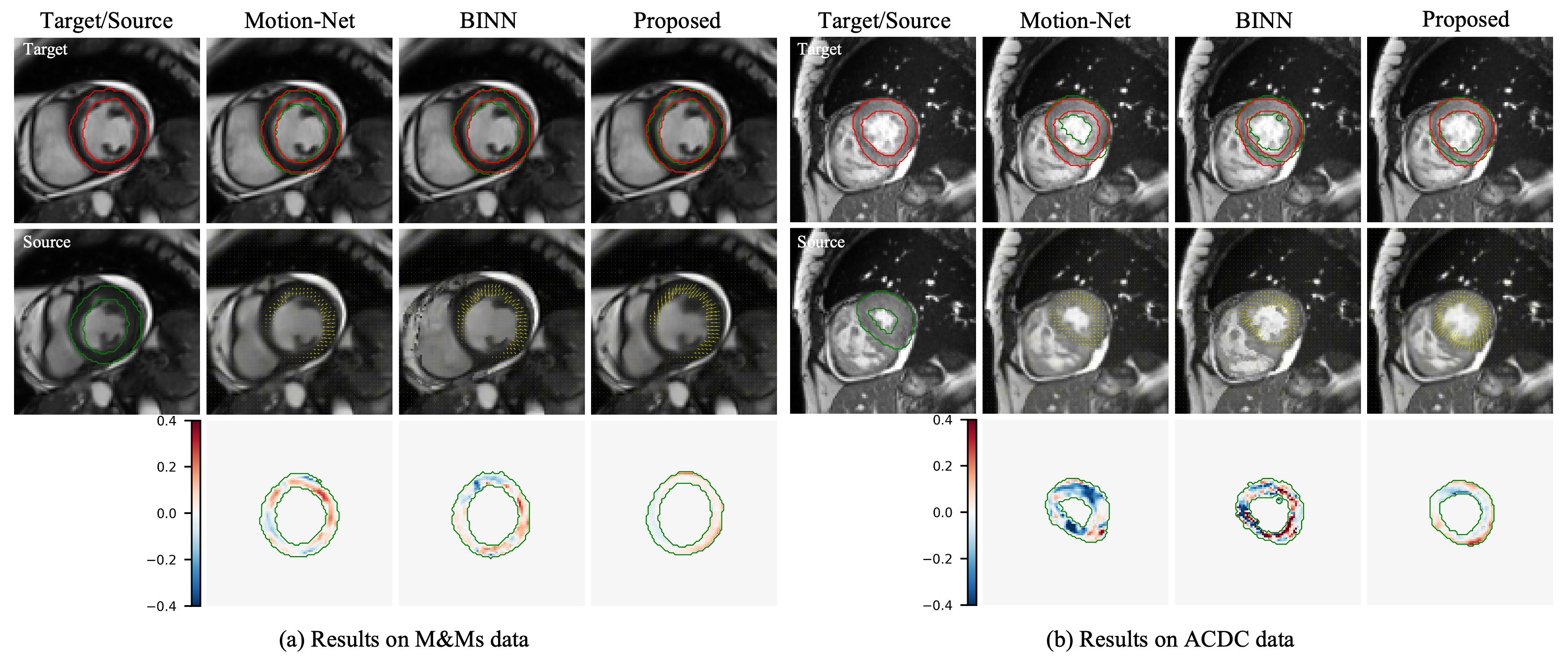}
\caption{Qualitative comparison results of Motion-Net, BINN and the Proposed approach on both (a) M\&Ms data and (b) ACDC data. The first column in both (a) and (b) shows the source and target image with contours of myocardial wall. Results in the first row show the warped source myocardial contour (green) in comparison to the target myocardial contour (red) overlaid on the target image. The second row presents the myocardial deformation fields in quiver plot (yellow arrows) overlaid on the transformed source image, in which BINN and Proposed regularize deformation only around myocardium. The last row provides the log of Jacobian determinant in myocardial regions.}
\label{fig:qualitative_rsults}
\end{figure*}

For qualitative results, here we visualise the estimated displacement fields, overlaps between target and warped LV myocardial contours, as well as the log of Jacobian determinant to show the deformation regularity, motion tracking accuracy and tissue incompressibility. The comparisons are shown in Fig. \ref{fig:qualitative_rsults}, where we mainly compare the DL-based approaches. The examples given are from both M\&Ms and ACDC datasets. From the first row in Fig. \ref{fig:qualitative_rsults}(a) and (b), we can observe that the proposed method achieved better motion tracking performance with a more accurate alignment of myocardial contours between warped source ES segmentation and target ED segmentation. This is particularly significant in comparison to Motion-Net which failed to generalize well to unseen data. BINN generalized better than Motion-Net in terms of the registration accuracy, however, in some cases this came with the sacrifice of the deformation regularity as shown in Fig. \ref{fig:qualitative_rsults}(b). This is mainly due to the reason that both of these methods are learning-based approaches, and they can still be affected by varying data distribution even with strong regularizations. In contrast, the proposed approach can maintain better deformation regularity and incompressibility in these examples and has been less affected by data shift issues relating to acquisition scanners and parameters as well as diseases.      
Though the incompressibility of the myocardium can not be guaranteed due to the out-of-plane motion on stacks of 2D slices, the results here still implies that the proposed data-driven prior informed by biomechanics can well parameterize the transformation model, which enables the generation of plausible deformation through latent space exploration. 
% A dynamic video showing the qualitative visualisations is also presented in supplementary material.

% Besides, the Jacobian determinant of deformation across the entire cardiac cycle on one subject is presented in Fig. \ref{fig:jacobian}. Here the proposed biomechanics-informed method is compared with the one using L2 norm, and it shows better preservation of the tissue volume at different cardiac phases, with Jacobian determinant being close to 1. 

\subsubsection{Strain analysis}
Furthermore, we also evaluated the performance of the proposed method on LV myocardial strain estimation. Table \ref{strain_evaluation} shows the comparison of peak radial strain (RR) and peak circumferential strain (CC) obtained from different methods on UKBB cine MRI data. 
The peak strain was also evaluated on apical, mid-ventricle and basal slices separately. To better understand the strains, we compared the predictions with reference values reported in \cite{bai2020population} {on the same group of subjects in the UKBB cohort with cine MRI.}
% The reference values were averaged across segments and genders from Supplementary Table 5 in \citep{bai2020population}, 
Motion tracking in \cite{bai2020population} was performed comprehensively by weighted averaging the forward and backward displacement fields derived via accumulation of successive inter-frame displacements using MIRTK toolkit. 
Compared with other methods, it can be observed from Table \ref{strain_evaluation} that the proposed approach can achieve closer and more consistent radial and circumferential strain value ranges with the reference values reported. To note that cine-derived radial strain is normally higher than tagged-derived ones \citep{ferdian2020fully}, and there is also a poor agreement in it even for commercial software packages \citep{cao2018comparison}. 
% Nevertheless, the proposed approach is able to achieve a more reasonable and accurate estimation of strains via motion tracking, which will be important for clinical analysis.
%

\begin{figure}[!t]
\centering
\includegraphics[width=\linewidth]{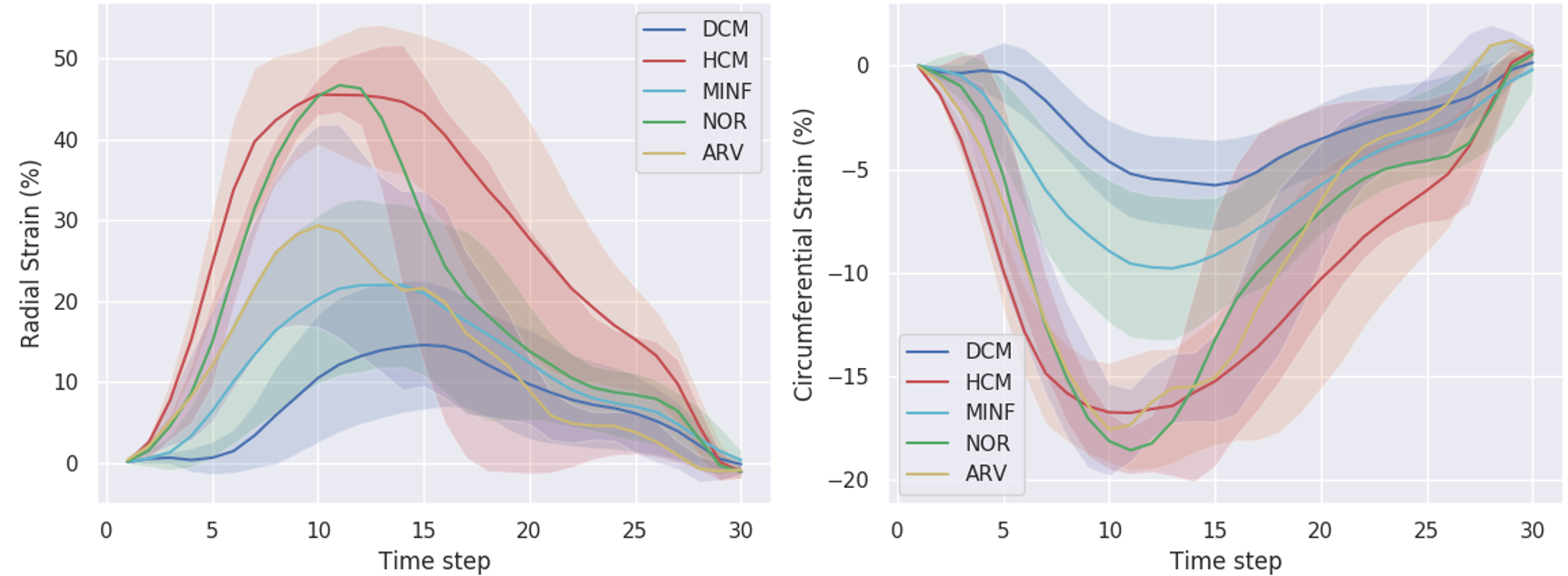}
\caption{Radial and circumferential LV myocardial strain curves estimated on ACDC data with four cardiovascular diseases (DCM, HCM, MINF, and ARV) and one normal group (NOR). }
\label{fig:strain_analysis}
\end{figure}

Besides, results of the global radial and circumferential myocardial strains across time is also presented in Fig. \ref{fig:strain_analysis}. Particularly, we plotted the strain curve on 44 ACDC data that consists of 30 time frames, and the curve shows the averaged results (solid lines) and their standard deviation (shaded areas) according to the groups of pathology, i.e., myocardial infarction (MINF), dilated cardiomyopathy (DCM), hypertrophic cardiomyopathy (HCM), abnormal right ventricle (ARV) and normal controls (NOR).
% with 10, 9, 8, 5 and 12 subjects respectively. 
From Fig. \ref{fig:strain_analysis}, we can observe that groups of DCM patients and MINF patients have reduced radial and circumferential peak strains compared to NOR, and the DCM strain curve also shows delayed time to peak myocardial strain. Besides, it can be seen that peak circumferential strain in HCM patients is also lower than that in NOR. These observations from our estimated strains are consistent with other clinical findings \citep{schultheiss2019dilated,hoit2011strain,nanda2013comprehensive,sun2019echocardiographic}, which indicates great potential of the proposed myocardial motion tracking approach for clinical interpretation and analysis.

\begin{table*}[!t]
  \centering
  \caption{Comparisons of peak strain values (\%) obtained from different methods at different slices. Reference values are obtained from cine MRI based on the same UK Biobank cohort \citep{bai2020population}. RR: peak radial strain; CC: peak circumferential strain. \textbf{Bold} numbers indicate results that are closest to the reference values.}
  \label{strain_evaluation}
\setlength{\tabcolsep}{3pt}
 \scalebox{0.9}
%  { \begin{tblr}{c|cccccccccc|cc}
 { \begin{tabular}{ccccccccccccc}
 \toprule
\multirow{2}*{Method} & \multicolumn{2}{c}{FFD-VP} & \multicolumn{2}{c}{dDemons}& \multicolumn{2}{c}{Motion-Net}&\multicolumn{2}{c}{BINN} &\multicolumn{2}{c}{Proposed} & \multicolumn{2}{c}{\textit{Reference}}\\
\cmidrule{2-13}
 & {RR} & {CC}   & RR  & CC & RR   & CC   & RR & CC   & RR & CC & \textit{RR} & \textit{CC}\\
  \midrule
  {Apical} & 16.9 & -10.1 & 30.0 & -14.2 & {26.5} & -14.1 & 32.5 & {-19.7} & \textbf{44.5} & \textbf{-22.1}  & 
%   \textit{51.0} & \textit{-26.9}
\textit{51.5} & \textit{-26.6}\\
  \midrule
 {Mid-ventricle} & 21.8 & -8.6 & 37.6 & -12.7 & 35.9 & -14.0 & {33.1} & {-18.1} & \textbf{50.0} & \textbf{-18.9}  & 
%  \textit{48.6} & \textit{-21.9} 
\textit{48.9} & \textit{-22.2}\\
 \midrule
Basal & 18.2 & -6.3 & 22.8 & -10.9 & 31.6 & -13.0 & {24.5} & {-16.4} & \textbf{47.0} & \textbf{-16.6} &
% \textit{45.6} & \textit{-22.6}
\textit{47.4} & \textit{-22.9}\\
    \bottomrule
%   \end{tblr}}
\end{tabular}}
\end{table*}

\subsubsection{Ablation study}
\begin{figure*}[!t]
\centering
\includegraphics[width=\linewidth]{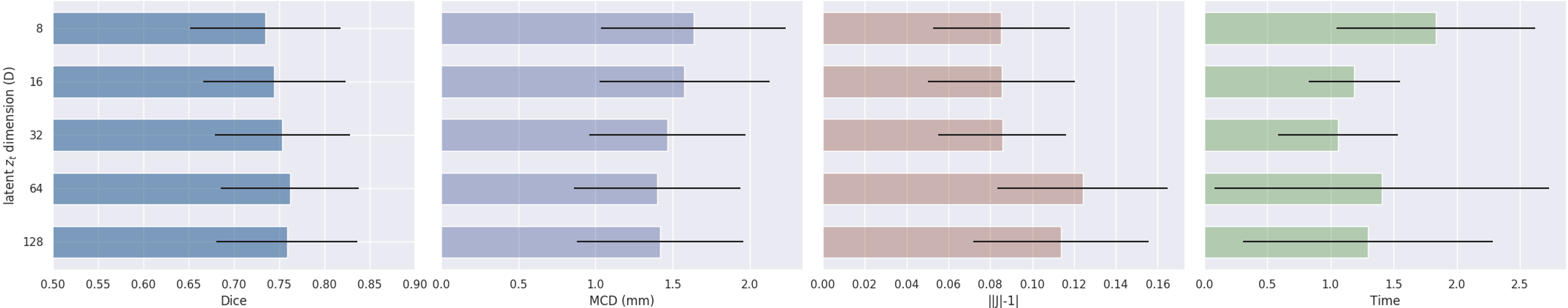}
\caption{{Ablation study on effects of the latent vector dimension $D$. Results are demonstrated on UKBB validation data with comparisons in terms of Dice, MCD, $||J|-1|$ and inference time.}}
\label{fig:ablation_z}
\end{figure*}

{ In this study, we performed ablation analysis on several components that constitute the proposed method. Particularly, we focus on investigating the effects of the latent vector dimension $D$, variation of parameter $\mu$ and the incorporation of temporal component.

\paragraph{Ablation on varying latent vector dimension}
\label{sec:varying_z}
We examine the effects of the latent vector ($\mathbf{z}_t$) dimension ($D$) by varying it on a set of values \{8, 16, 32, 64, 128\} and comparing their resulting performance on the UKBB validation set in terms of registration accuracy, volume preservation and inference time. The results are shown in Fig. \ref{fig:ablation_z}, where the mean and standard deviations are given. It can be seen that as $D$ increases, the accuracy of the registration (Dice and MCD) gets improved, mainly due to the richer representation of the data and the increased number of parameters to be optimized. However, this leads to a reduction on the volume preservation which could be possibly resulted from the increased complexity in manifold traversing. Besides, the test-time registration speed depends both on the dimension of the optimized latent vector and the number of iterations needed for convergence. From Fig. \ref{fig:ablation_z}, we can observe that $D=32$ achieves the optimal balance between these two factors with the lowest inference time, whereas lower dimension ($D=8$ and $D=16$) would require more iterations to converge and higher dimension needs longer forward and backward propagation time. Considering these different aspects of performances, $D=32$ can achieve a good trade-off and serves as the best choice here. 
}

\begin{figure}[!t]
\centering
\includegraphics[width=\linewidth]{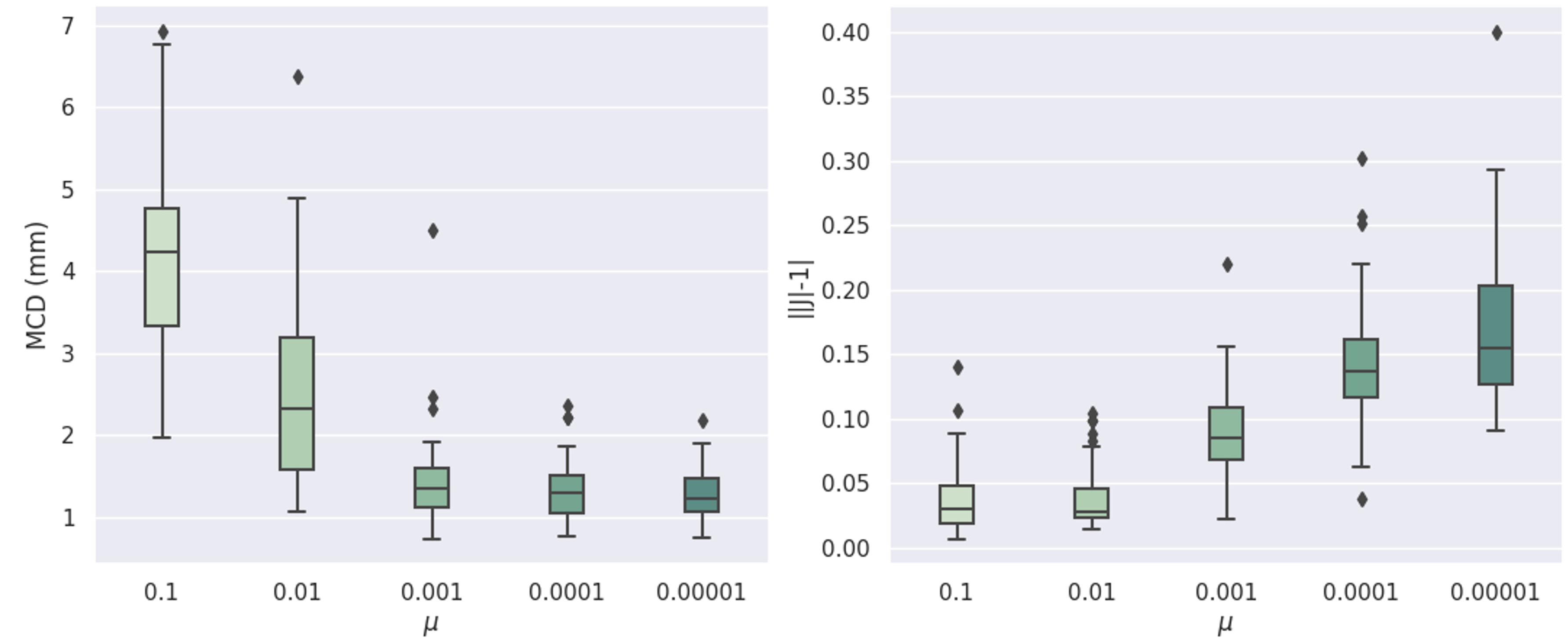}
\caption{{Ablation study on effects of parameter $\mu$. Results are compared in terms of MCD and $||J|-1|$ on different values of $\mu$.} }
\label{fig:ablation_mu}
\end{figure}

{
\paragraph{Ablation on varying $\mu$}
We have also explored the effects of $\mu$ in Eq. \ref{L_sim_l2} on the motion tracking performance with the UKBB validation set. Here $\mu$ was varied with values of \{0.1, 0.01, 0.001, 0.0001, 0.00001\}, and the performance was evaluated using quantitative measures of MCD and $||J|-1|$, as shown in Fig. \ref{fig:ablation_mu}. Since $\mu$ balances between the data fidelity and the likelihood of falling on the learnt manifold, we expect that as $\mu$ decreases, the registration accuracy would be improved whereas the incompressiblity attribute would get weakened. This in fact aligns with our observation in Fig. \ref{fig:ablation_mu}. Similarly, $\mu=0.001$ shows a better trade-off between these two measures compared to other values, producing satisfactory registration accuracy with a reasonable amount of tissue incompressibility. This justifies our choice of $\mu$ used in previous experiments.
}

\begin{figure*}[!t]
\centering
\includegraphics[width=.9\linewidth]{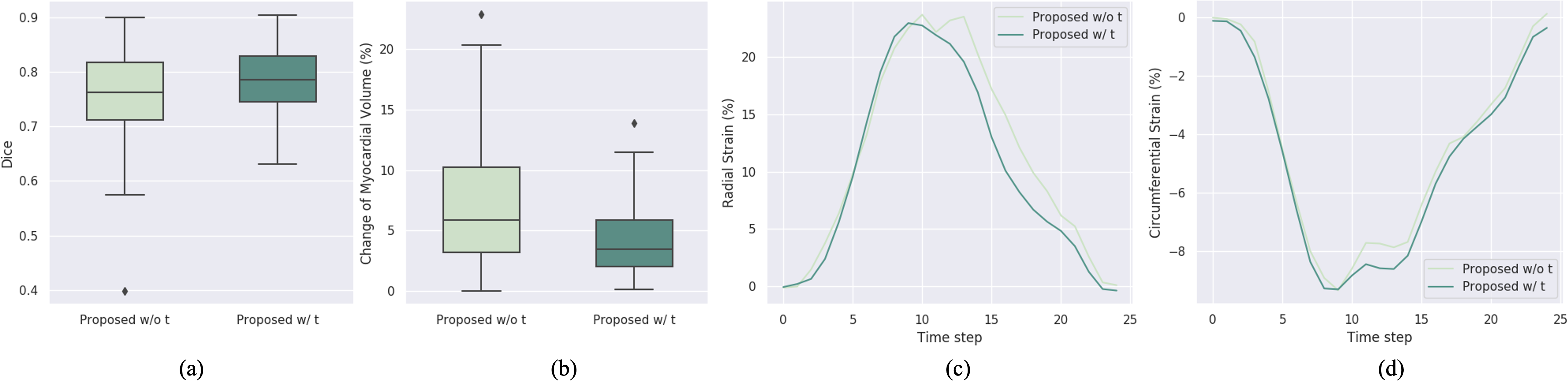}
\caption{Ablation study on effects of the incorporation of temporal information for LV myocardial motion tracking. Results are demonstrated on M\&Ms data with comparisons in terms of (a) Dice, (b) change of myocardial volume, (c) radial strain and (d) circumferential strain. }
\label{fig:ablation}
\end{figure*}

\paragraph{Effect of temporal component}
In addition, we also study the effects of introducing temporal information for the motion tracking performance. Particularly, we compared the proposed approach that takes temporal sequences into account (Proposed w/ t) with the one that only considers independent pairs of frames for both manifold learning and registration (Proposed w/o t). To enhance the latent capacity of Proposed w/o t to encode the deformation, we increase the dimension of $\mathbf{z}_t$ to 256 for Proposed w/o t. The ablation results are shown in Fig. \ref{fig:ablation}, where we compared the myocardial motion tracking performance in terms of Dice and volume preservation between ES and ED frames on volumes of M\&Ms data, as well as the radial and circumferential strains across cardiac cycle. It can be observed that when taking account of the temporal information, the proposed approach achieved better myocardial volume preservation as well as a higher Dice score than Proposed w/o t. The estimated strains shown in Fig. \ref{fig:ablation}(c) and (d) also present smoother strain values along time in comparison to Proposed w/o t. This indicates that the incorporation of temporal information allows the learnt manifold to potentially capture temporal behaviours of the simulations, which could represent more realistic motions than those without temporal information. The improvement is also likely due to the reason that temporal information can serve as an additional constraint for latent space exploration, which enforces the search to fall on the biomechanics-plausible manifold with the consideration of temporal consistency.

\section{Discussion}
\label{sec:discussion}
In this work, we have demonstrated that the proposed method is capable of achieving competitive myocardial motion tracking performance with desired properties and generalizability. Different from existing image registration approaches, we integrated a data-driven application-specific prior based on a generative neural network with an optimization model-based registration via latent space exploration. Compared with learning-based approaches trained on specific datasets, the proposed method presents better robustness and generalizability to imaging data distribution shifts, as shown in Table \ref{motion_evaluation} and Fig. \ref{fig:qualitative_rsults}. This is mainly due to three inherent natures of the proposed approach: first, the proposed method is imaging domain-invariant which does not encode any imaging information, and thus it will not be susceptible to imaging data distribution shifts; second, the proposed biomechanical constraint either in explicit or implicit forms is able to regularize the generated deformations to be plausible, and therefore results in its less sensitivity to domain shifts in learning settings; additionally, the latent space exploration on the biomechanically plausible manifold offers the flexibility in searching for the optimal transformations for the target sequences during inference, {which thus also allows the generalization from healthy to diseased subjects (Fig. \ref{fig:strain_analysis})}. Besides, the data-driven prior is also advantageous over the explicit generic regularization terms (e.g. smoothness penalty), as it could potentially capture realistic and complex properties from observational data that could not be explicitly specified. The resulting generative transformation model thus is able to embed the prior knowledge into the parameterized transformation network, which also helps to avoid the explicit regularization as in standard methods.
% hyper-parameter tuning in Eq. \ref{objective_function} .

In terms of the myocardial motion tracking speed, learning-based approaches are superior as they do not need to perform any update at inference time. In contrast, the proposed method relies on optimization in latent space, which would require relatively more time to update the latent motion matrix. Given the optimization technique we employed, we found that the convergence can be achieved within 60 iterations on UKBB data and within 300 iterations on more challenging data (ACDC and M\&Ms data). The myocardial motion tracking time for UKBB sequence data with 50 frames per slice is an average of 1.08s compared to 0.08s in BINN and Motion-Net, {and is an average of 2.96s on ACDC and M\&Ms datasets}. Considering the performance gain of the proposed method, the slightly longer motion tracking time can be acceptable in practice. In addition, there were also some other approaches that proposed iterative refinement over the entire network parameters at test time \citep{lee2019image}. However, these works are more computationally expensive as more parameters require updates during inference, whereas our proposed approach only needs to update the latent vectors and thus is more efficient. Besides, though refining network parameters for Motion-Net or BINN at test time would possibly lead to better motion tracking accuracy, they could be at the risk of violating the desired biomechanical properties during refinement.

\begin{figure}[!t]
\centering
\includegraphics[width=.8\linewidth]{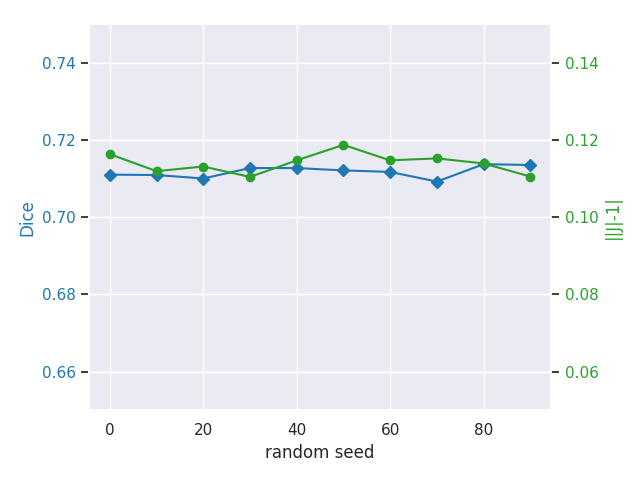}
\caption{{Performance on a single ACDC subject with different random initializations of latent motion matrix $\mathbf{z}$.} }
\label{fig:random_init}
\end{figure}

With respect to the manifold learning, we can in fact add more constraints on the reconstruction terms in Eq. \ref{eq:VAE} to enforce it to learn more of the desired properties from the biomechanical simulations, which could potentially facilitate the learning and capture more diverse patterns. For latent space exploration in Eq. \ref{L_sim}, different initializations of the latent motion matrix could lead to results converging to different optimum, however, we empirically found that the averaged performance remained similar on the cohort. { An example of this is shown in Fig. \ref{fig:random_init}, where different random initializations of the latent motion matrix $\mathbf{z}$ were performed on a single subject and the results were found to be relatively stable with respect to different random seeds.} More advanced optimization and initialization techniques could also be investigated for the manifold traversing, such as spherical optimizer as proposed in \citep{menon2020pulse}. These could possibly further improve the motion tracking performances, and we will leave these for future explorations. In addition, {though the present method shows the motion tracking on short axis views, there is also possibility to generalize the method to long axis views and we will investigate this in future work. Besides, one limitation of} our current model is that it performs on 2D myocardial motion tracking, {where the cardiac deformation is constrained to be in-plane}. This is mainly limited by the existing available data where there are only 2D acquisitions of cardiac cine MRI. Nevertheless, this is consistent with the clinical practice in evaluating the cardiac motion and strains on 2D tagged MRI. For future work, we will extend this work to 3D applications, where we will generate 3D simulations and predict deformation fields based on super-resolved 3D cardiac images, {to consider more realistic and complex cardiac motion. Possible challenges of extending this work to 3D may include the increased computational burden due to both 3D data and 3D networks, as well as the prior learning in capturing more complicated through-plane cardiac motion.} We employed simplified biomechanical simulation scheme in this work and it can be further improved in future work to reflect the in-vivo physiological condition, e.g., more realistic constitutive relations. Besides, it will also be interesting to investigate the applicability of the proposed approach to more general scenarios, such as to other anatomical structures and modalities.

\section{Conclusion}
In this paper, we have presented a novel biomechanics-informed generative model for myocardial motion tracking in cardiac cine MRI with latent space exploration. A temporal VAE-based model is proposed to learn a manifold for biomechanically simulated deformations and to parameterize such prior knowledge in a deep generative neural network. A learning-based generative transformation model can therefore be derived, where the deformations are  compactly represented in the low dimensional latent space. Myocardial motion tracking then can be subsequently performed by traversing the biomechanically plausible deformation manifold to search for the optimal transformations given the cardiac sequences. Experimental results have shown that the proposed method outperforms the other competing approaches, achieving better motion tracking accuracy with more reasonable myocardial motion and strain estimates. It also enables better generalizability to imaging data distribution shifts, which is promising for its future practical application in clinical use. 
% For future work, we will extend the method for 3D motion tracking and investigate on incorporating physiology modelling.

\section*{Acknowledgement}
This work was supported by EPSRC programme grant SmartHeart (EP/P001009/1) and EPSRC grant DeepGeM (EP/W01842X/1).
%%Harvard
\bibliographystyle{model2-names.bst}\biboptions{authoryear}
\bibliography{refs}

% \section*{Supplementary Material}

% Supplementary material that may be helpful in the review process should
% be prepared and provided as a separate electronic file. That file can
% then be transformed into PDF format and submitted along with the
% manuscript and graphic files to the appropriate editorial office.

\end{document}